\documentclass[nohyper,12pt,letterpaper]{JHEP3}
\usepackage{epsfig}
\usepackage{amsmath}

\DeclareSymbolFont{AMSa}{U}{msa}{m}{n}
\DeclareSymbolFont{AMSb}{U}{msb}{m}{n}
\let\Box\relax
\DeclareMathSymbol{\Box}{\mathord}{AMSa}{"03}

\def \IR{{\mathbb R}}
\def \IC{{\mathbb C}}

\def\odots{\mathinner{\mkern1mu\raise0pt\vbox{\kern7pt\hbox{.}}
\mkern2mu\raise4pt\hbox{.}\mkern2mu\raise8pt\hbox{.}\mkern1mu}}

\def\sgn{{\textrm{sgn}}}
\title{Finite Number of States, de Sitter Space and Quantum Groups at Roots of Unity}
\author{Philippe Pouliot \\
\begin{tabular}{lcl}
on leave from &&Physics Department\\
Physics Department & \& &Queen Mary, University of London\\
University of Texas at Austin & & London, E1 4NS, UK\\
Austin, TX 78712 USA  && \email{p.pouliot@qmul.ac.uk}
\end{tabular}}

\abstract{This paper explores the use of a deformation by a root
of unity as a tool to build models with a finite number of states
for applications to quantum gravity. The initial motivation for
this work was cosmological breaking of supersymmetry. We explain
why the project was unsuccessful. What is left are some
observations on supersymmetry for q-bosons, an analogy between
black holes in de Sitter and properties of quantum groups, and an
observation on a noncommutative quantum mechanics model with two
degrees of freedom, depending on one parameter. When this
parameter is positive, the spectrum has a finite number of states;
when it is negative or zero, the spectrum has an infinite number
of states. This exhibits a desirable feature of quantum physics in
de Sitter space, albeit in a very simple, non-gravitational
context.}


\preprint{UTTG-12-01, QMUL-PH-03-06}

\begin{document}
\section{Introduction}
We know very little about quantum gravity. Perhaps the most
important insight about what a theory of quantum theory of gravity
ought to look like is the holographic principle~\cite{'tHooft:gx}.
This already suggests a drastic reduction in the number of degrees
of freedom required for a theory of quantum gravity, compared to
what we have in a quantum field theory~\footnote{This is
particularly true in loop quantum gravity, where quantum groups at
roots of unity and finite/discrete number of states naturally
enter the formalism. See~\cite{Crane:vv} and, for an extensive
list of references to the original literature, the excellent
recent article~\cite{Smolin:2002sz}.}. The observation by
astronomers that the expansion of the Universe is accelerating
could be explained by a non-zero dark energy or cosmological
constant and that we live in an asymptotically de Sitter
space-time. According to Fischler~\cite{fischler}, to
Banks~\cite{Banks:2000fe}, to Bousso~\cite{Bousso:2000nf}, and to
others, this suggests that a further, even more drastic, reduction
is required. Namely, that only a finite number of quantum states
is sufficient to describe all the causal physics that happens
within one observer's cosmological horizon.

This would indicate that there is something fundamentally wrong
with the quantum field theories and string theories that we
usually work with: they all have an infinite number of states.

Furthermore, a cosmological constant necessarily breaks
supersymmetry, and Banks conjectured~\cite{Banks:2000fe} that the
scale of supersymmetry breaking is directly tied to the size of
the cosmological constant. Note however that the scale associated
with supersymmetry breaking~$m_{SUSY}$ is much larger than that
set by the cosmological constant $\Lambda$. Thus, supersymmetry
has to be broken by a ``large'' amount if the origin of
supersymmetry breaking is the cosmological constant.
Phenomenologically, a relation such as $m_{SUSY}\propto
\Lambda^{1/8}$ would work.

In this paper, we are looking for toy models that would support
this conjecture. There are of course many ways to truncate
theories to a finite number of states. The main tool that we will
be using is deforming by a root of unity the simple harmonic
oscillators that are ubiquitous in perturbation theory. In this
way, we readily obtain theories with a finite number of states.
However, this procedure does not necessarily break supersymmetry.
We have not found an example where supersymmetry is broken by a
``large'' amount, but we have not ruled out their existence
either. Thus our approach has been unsuccessful so far. Also, the
idea of using quantum groups as a discretization tool is not new,
see for example~\cite{Majid:we} or~\cite{Major:1995yz}. We hope
however that our presentation is new, emphasizing the simplicity
of using $q$-bosons and quantum groups.

In summary, in section 2, we review the properties of $q$-bosons,
and discuss their supersymmetric properties. In section 3, we
review $SU(2)_q$ and its thermodynamics properties and draw a
caricature of how it could be applied to black hole physics. In
section 4, we review the non-commutative quantum mechanics model
of Bellucci, Nersessian and Sochichiu and supersymmetrize it.

\section{The $q$-deformed oscillator}
\subsection{Review of the $q$-boson algebra for $q$ real}
Although perhaps not as familiar as the quantum group $SU(2)_q$,
we will begin with $q$-bosons. This is a simpler starting point,
since there is only one degree of freedom. In addition, all
quantum groups can be built out of $q$-bosons. Please see the
excellent book~\cite{Biedenharn:vv} for further details and an
extensive list of references to the original literature.

There is an essential distinction between $q$ real and $q$ a root
of unity. Although our true interest is in the case where $q$ is a
root of unity, let us discuss first the simpler case when $q$ is a
positive real number. When $q\in \IR^+$, the representations of
the $q$-boson algebra are rather similar to the ordinary boson.
When $q$ is real, there is an infinite number of states and when
$q$ is a root of unity, there will be a finite number of states.

The $q$-boson algebra for $q$ real consists of the creation and
annihilation operators $a_+$ and $a_-$, and the number operator
$N$~\footnote{Our conventions differ from~\cite{Biedenharn:vv} by
$q^{1/2}\rightarrow q$, $a\rightarrow a_+$ and $\bar a \rightarrow
a_-$.}. They satisfy the defining relations:
\begin{equation}
a_- a_+ - q a_+  a_- = q^{-N}, \qquad [N,a_{\pm}]=\pm a_{\pm}.
\label{qrealharmonic}
\end{equation}
The last relation can also be written as $q^Na_\pm q^{-N}=q^{\pm
1} a_\pm$. We see that these relations interpolate between boson
($q=1$) and fermion ($q=-1$) commutation relations, which also
makes it clear that working out the details of the $q$-fermion is
easy (see Hayashi in~\cite{Biedenharn:1989jw}). There are of
course many equivalent ways of writing
equation~\ref{qrealharmonic}. One common one is $A_+=a_+q^{N/2}$,
$A_-= q^{N/2}a_-$ for which the first relation
in~\ref{qrealharmonic} becomes $A_-A_+-q^2A_+A_-=1$~\footnote{They
are not quite equivalent, as explained in~\cite{Burban:1993},
where three different deformations are considered when $q$ is
real. The main physical distinction between the cases is that the
operators of position and momentum can be bounded or unbounded.
When unbounded, these operators turn out to not be self-adjoint
but to have a self-adjoint extension. This is the case for
equation~\ref{qrealharmonic}, which is the second case considered
in~\cite{Burban:1993}. In all cases, for $q$ real, their spectrum
is continuous. For $q$ a root of unity, the spectrum is finite and
therefore the issue does not arise.}.

This algebra is known as the $q$-Heisenberg-Weyl algebra
$U_q(h_4)$ but we will call it here the $q$-boson algebra for
short. For doing quantum mechanics, one needs a hermitian
conjugation (an involution) $\dagger$; here it is simply
\begin{equation}
a^\dagger_- = a_+,\qquad a^\dagger_+ = a_-, \qquad N^\dagger = N.
\label{qrealhermiticity}
\end{equation}
The notion of a $q$-integer will be useful:
\begin{equation}
[x] = \frac{q^{x}-q^{-x}}{q-q^{-1}}.
\end{equation}
The algebra~\ref{qrealharmonic} has a central element:
$q^{-N}([N]-a_+a_-)$. The actual algebra we are interested in is
equations~\ref{qrealharmonic} modded out by setting this central
element to zero~\cite{Petersen:1994nm}. The result is that there
is only one unitary irreducible representation of this restricted
algebra, and it is infinite dimensional. It is the one we would
expect, closely related to that of the ordinary boson. This
representation can be realized by a Fock space, with a vacuum
state $|0\rangle$ that is annihilated by $a_-$, and
\begin{equation}
a_+|n\rangle = \frac{a_+^n}{\sqrt{[n]!}}|0\rangle, \qquad
N|n\rangle = n|n\rangle, \label{qharmonicrep}
\end{equation}
with the $q$-factorial defined recursively by $[n]!=[n] ([n-1]!)$.
In the restricted algebra, the first relation
in~\ref{qrealharmonic} is actually equivalent to the two relations
\begin{equation}
a_+a_- = [N], \qquad a_-a_+=[N+1].
 \end{equation}

The position and momentum operators are defined in the usual way:
\begin{equation}
X= \frac{a_++a_-}{\sqrt{2}}\qquad {\textrm {and}}\qquad P=
\frac{-a_++a_-}{i\sqrt{2}}.
\end{equation}
They obey the commutation relation
\begin{equation}
[X,P] = i([N+1]-[N]).
\end{equation}
In physical terms, this means that Planck's constant is no longer
a constant, but depends on the excitation energy of the system,
and actually diverges at very high energy. Readers are referred to
Biedenharn in~\cite{Biedenharn:1989jw} for other such statements
on the physics of $q$-bosons.

For the hamiltonian of a $q$-deformed oscillator, we make the
natural choice
\begin{equation}
H= \frac{1}{2}\left(a_+a_-+a_-a_+\right) =\frac{1}{2}([N]+[N+1]).
\end{equation}
Its spectrum for small values of $n$ is almost evenly spaced, and
thus very similar to the ordinary simple harmonic oscillator (see
first two columns of figure 3). For large values of $n$ however,
for $q<1$, $[n]\rightarrow q^{-n}$ and grows exponentially fast
and the spectra become very different.

\subsection{Review of the $q$-boson algebra for $q$ a root of unity}

Let us now describe the $q$-boson for $q$ a root of unity. Our
presentation will be elementary and our focus is on illustrating
the finiteness of the number of states. The algebra consists of
the creation and annihilation operators $a_+$ and $a_-$, and
of~$L$. The defining relations are~\cite{Biedenharn:vv} (page
190):
\begin{equation}
a_- a_+ - q a_+  a_- = L^{-1}\quad \textrm{and} \quad La_\pm
L^{-1}= q^{\pm 1}a_{\pm}, \label{qharmonic}
\end{equation}
for $q
= e^{2 \pi i/k}$ a root of unity. The number operator $N$, such
that $L=q^{N}$, is not well-defined: it is multi-valued and
periodic; however, operators such as $[N]$ are well-defined and
can be expressed in terms of $L$ and $q$. It follows from
\ref{qharmonic} that
\begin{equation}
a_+ a_- = \frac{L-L^{-1}}{q-q^{-1}} = [N] \quad \textrm{and}\quad
a_- a_+ = \frac{q L-q^{-1}L^{-1}}{q-q^{-1}} = [N+1] .
\end{equation}
We have $[x]=\frac{q^x-q^{-x}}{q-q^{-1}}=\frac{\sin{2\pi
x/k}}{\sin{2 \pi/k}}$, so that now there are identities that
follow from properties of the sine function:
$[x+\frac{k}{2}]=-[x]$, $[x+k]=[x]$,$[\frac{k}{2}]=0$. With these
relations, one can show~\cite{Biedenharn:vv} by induction that
$a_-a_+^n=q^n a_+^n a_- + [n]  a_+^{n-1}L^{-1}$. Specializing to
$k$ odd for now, $a_\pm^k$ commutes with $a_\pm$ and $L$. Because
an operator like $a_+a_-$ can take both positive and negative
values, one has to define the hermitian conjugation $\dagger$ as
follows in order to fix up the signs: $a_+^\dagger = a_-
\epsilon([N])$, $a_-^\dagger = \epsilon([N]) a_+$, and $L^\dagger
= L^{-1}$, where $\epsilon$ is a sign function: $\epsilon(x) =
\left\{
\begin{aligned}
1,&\qquad \textrm{if}\ x\ge 0,\\
-1,&\qquad \textrm{if}\ x<0.
\end{aligned}
\right.$

We can now review the representation theory~\cite{Biedenharn:vv}.
For $q$ an odd root of unity, it is much more complicated than for
$q$ real. There are nilpotent, cyclic, semi-cyclic and
indecomposable representations. However, only one of them is
interesting to us.

We are interested in the representation for which $a_\pm^k=0$;
hence it is called nilpotent. It is unitary and irreducible. Label
its carrier space by $|n\rangle$, $n=0,1,\ldots,k-1$. Then the
generators act by:
\begin{align}
L\, |n\rangle & = q^n\, |n\rangle, \qquad a_+\, |k-1\rangle  = 0,
\qquad a_-\, |0\rangle  = 0 \nonumber \\
a_+\, |n\rangle & = \sqrt{|[n+1]|}\, |n+1\rangle, \qquad n \neq k-1, \\
a_-\, |n\rangle & = \epsilon([n]) \sqrt{|[n]|}\, |n-1\rangle,
\qquad n \neq 0. \nonumber
\end{align}
Thus it can be built by acting in the usual way with creation
operators: $|n\rangle = \frac{(a_+)^n}{\sqrt{|[n]|!}}|0\rangle$,
with $a_-|0\rangle=0$ and $|k\rangle=0$, and $|\ |$ denotes the
absolute value. So the representation has a finite number of
states, $k$~\footnote{For $k$ even, there are two nilpotent
representations, with $a_\pm^{k/2}=0$, and each with $k/2$
states.}.

There are more representations, those for which $a_\pm^k\neq 0$.
We can simply ignore them for our purposes\footnote{In a continuum
quantum field theory, these representations would be solitons.
They are characterized by a continuous parameter $\phi$, defined
modulo $2\pi$, analogous to an (additive) magnetic charge. In
quantum field theory, because of unitarity, they would necessarily
arise. They would be created  non-perturbatively in pairs of
``charge'' $\phi$ and $-\phi$ of total charge zero, and ignoring
them would violate unitarity. For our discretized theories with a
finite number of states, the rules are unitary without these
solitons. For completeness, we will review now these cyclic and
semicyclic representations, for odd $k$. They are labelled by
$l\in \IC^*$, $\mu$, $\nu\in \IC$. The action on the carrier space
$|n\rangle$, $n=0,1,\ldots,k-1$ is
\begin{align*}
L\, |n\rangle & = l q^n\, |n\rangle, \qquad a_+\, |k-1\rangle  =
\mu |0\rangle,
\qquad a_-\, |0\rangle  = \nu |k-1\rangle \\
a_+\, |n\rangle & = |n+1\rangle, \qquad n \neq k-1, \\
a_-\, |n\rangle & = (l^{-1}[n] + q^n\mu\nu)\, |n-1\rangle, \qquad
n \neq 0,
\end{align*}
and the values of the Casimirs are $L^k=l^k$, $a_+^k = \mu$,
$a_-^k = \nu\prod_{n=1}^{k-1}(l^{-1}[n] + q^n \mu\nu)$, related by
$a_-^k a_+^k = [N+1][N+2]\cdots [N+k]$. On the restricted algebra,
defining an angle $l=e^{i\phi}$, $-\pi\le \phi < \pi$, there is a
further relation $\mu\nu=\frac{l-l^{-1}}{q-q^{-1}}=[\phi]$ and the
action of $a_-$ simplifies to $a_-|n\rangle =\left(
\frac{q^nl-q^{-n}l^{-1}}{q-q^{-1}}\right) |n-1\rangle $.}.

Now a few words about the properties of the position and momentum
operators. We can define the manifestly hermitian
coordinates~\footnote{We could also have used the hermitian
operator $\tilde X = \frac{1}{\sqrt 2}(a_-+a_-^\dagger)$ (or
$\tilde P =\frac{1}{i\sqrt 2}(a_-+a_-^\dagger)$). $\tilde X$
almost commutes with $X$ (only 2 entries are non-zero in the
matrix of the commutator $[X,\tilde X]$), and their eigenvalues
are the same. However, $\tilde X$ does not have a simple
commutation relation with $P$).}
\begin{equation}
X = \frac{1}{\sqrt 2}(a_++a_+^\dagger)\qquad \textrm{and} \qquad P
= \frac{1}{i\sqrt 2}(a_+-a_+^\dagger).
\end{equation}
 They satisfy
\begin{equation}
[X,P]=i\left(|[N+1]|-|[N]|\right).
\end{equation}
We have studied the spectrum numerically. In figure 1, we can find
the discrete, finite, spectrum of $X$ for $q=e^{2\pi i/25}$. The
total length of the line grows like $k^{1/2}$. $X=0$ is always an
eigenvalue, and the spectrum is symmetric with respect to
$X\rightarrow -X$. Except close to the origin, we see that two
eigenvalues reside at each value of $X$, with a very narrow
splitting of order $e^{-c k}$, where $c$ is independent of $k$ but
grows larger for eigenvalues farther from the origin. Finally,
such pairs of very nearly degenerate eigenvalues are separated
from their next neighbors by a distance that shrinks like
$c'k^{-1/2}+{\cal O}(k^{-3/2})$ where $c'$ is another constant.
Thus, we see that a $q$-deformation is a subjectively nice
discretization of the real line, in the sense that even for
relatively small $k$, the line is chopped into $k$ bits all of
much the same length. \vbox{\begin{center}
\epsfig{file=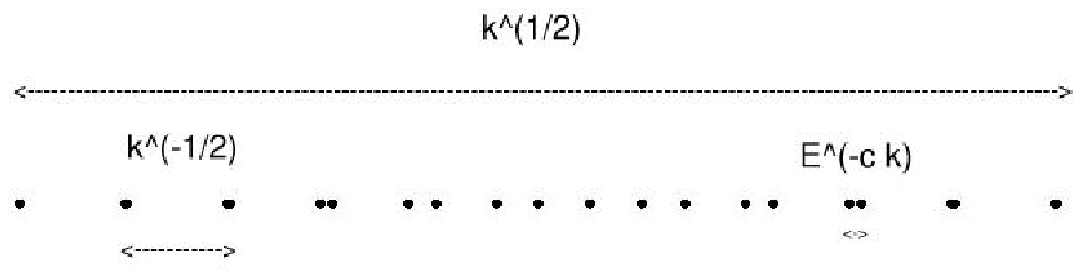}
\end{center}
\begin{center}
Figure 1. The discretization of the spectrum of $X$.
\end{center}}

\subsection{Supersymmetry of the $q$-oscillator at an odd root of unity}
In this section, we make a very simple, but strange, observation:
the ordinary, non-supersymmetric, simple harmonic oscillator can
lead to a supersymmetric model after $q$-deformation by an odd
root of unity.

There are at least 3 hermitian hamiltonians one can consider, that
reduce to the usual s.h.o. in the limit $q\rightarrow
1$~\footnote{In the case where $k$ is an even root of unity, the
hamiltonians $H_A$, $H_B$ and $H_C$ coincide and the spectrum is
positive and exactly doubly degenerate.}:
\begin{align}
H_A&= \frac{1}{2}(a_+a_-+a_-a_+), \qquad
H_B= \frac{1}{2}|a_+a_-+a_-a_+|, {\textrm {and}}\nonumber \\
H_C&=
\frac{1}{2}\left(|a_-a_+|+|a_+a_-|\right)=\frac{1}{2}(a_+^\dagger
a_++a_-^\dagger a_-)=\frac{1}{2}(|[N+1]|+|[N]|).
\end{align}
Their spectra, which consist of $k$ states, are compared in figure
2, along with the undeformed s.h.o. The maximum energy is $E_{max}
= \frac{1}{\sin(\frac{2\pi}{k})} \rightarrow \frac{k}{2\pi}$, and
it grows linearly at large $k$ for all three hamiltonians.

The spectrum of hamiltonian $H_A$ has both negative and positive
energies, and is symmetric with respect to the zero energy point.
It is like a Dirac sea (despite there being no fermions here.) The
spectrum of $H_C$ has doubly degenerate states, except for one
state of energy very close to the two ground states. This state
has energy $\frac{1}{2\cos\frac{\pi}{k}}\rightarrow
\frac{1}{2}+\frac{\pi^2}{4k^2}$. As $k\rightarrow \infty$, $H_C$
has three ground states of energy $1/2$. \vskip1cm
\vbox{\begin{center} \epsfig{file= 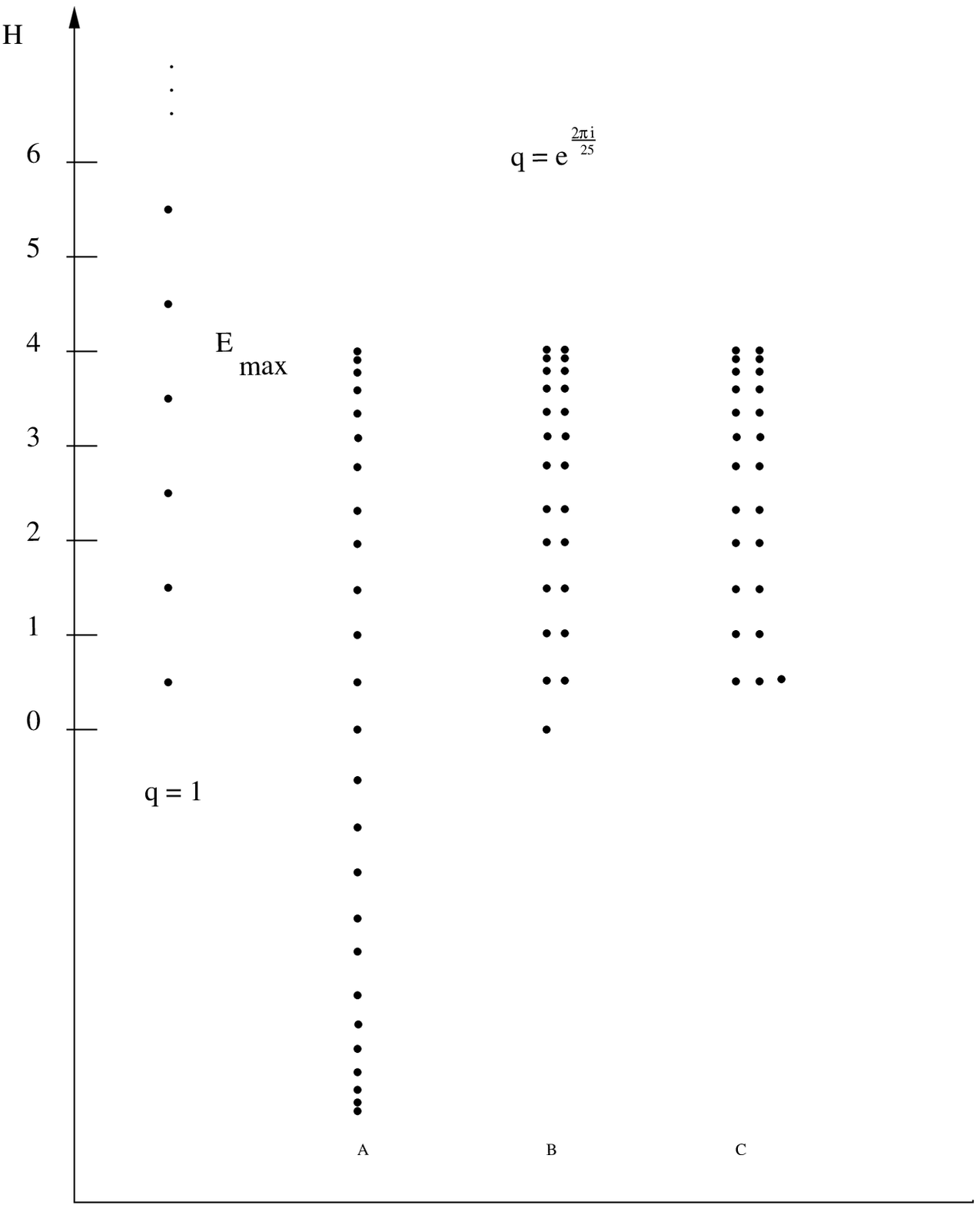}
\end{center}
\begin{center}
Figure 2. The spectra of the ordinary s.h.o. along with $H_A$,
$H_B$ and $H_C$.
\end{center}}\vskip1cm
The spectrum of $H_B$ is supersymmetric. This is rather strange:
never have we mentioned fermions, yet this hamiltonian has a
single zero energy ground state, and the rest of its spectrum is
doubly degenerate. For concreteness, one can write down two
hermitian supercharges:
\begin{equation}
Q_1 = R\, \Sigma_1\quad \textrm{and}\quad Q_2 = R\, \Sigma_2,
\end{equation}
where
\begin{equation}
\Sigma_1 = \left( \begin{smallmatrix}
&&&&&& 1 \\
&&&&& \odots \\
&&&& 1 \\
&&& 0 \\
&& 1 \\
& \odots \\
1\\
\end{smallmatrix}\right),
\qquad \Sigma_2 = \left( \begin{smallmatrix}
&&&&&& -i \\
&&&&& \odots \\
&&&& -i \\
&&& 0 \\
&& i \\
& \odots \\
i\\
\end{smallmatrix} \right).
\end{equation}
and $R$ is the diagonal matrix of entries
$\frac{1}{\sqrt2}\sqrt{|[N+1]+[N]|}$. They satisfy
$\{Q_\alpha,Q_\beta\}= 2\delta_{\alpha\beta} H_B$ for
$\alpha,\beta=1,2$. We hope that this $q$-boson with hamiltonian
$H_B$ could be an interesting way to introduce fermions and
supersymmetry in some applications.

\subsection{More supersymmetric examples}
When $q$ is real $\in(0,1)$, a straightforward way to
supersymmetrize the $q$-deformed oscillator is by introducing two
ordinary fermions and two hermitian conjugate supercharges:
\begin{equation}
Q_+=\sqrt{2}(a_-\psi_+),\quad\textrm{and}\quad
Q_-=\sqrt{2}(a_+\psi_-),
\end{equation}
with $\{\psi_\pm, \psi_\pm\}=1$, $\{\psi_\pm,\psi_\mp\}=0$,
$\psi^\dagger_\pm=\psi_\mp$ and $a^\dagger_\pm= a_\mp$. The
hamiltonian is:
\begin{equation}
H=\frac{1}{2}\{Q_+,Q_-\}=[N]+([N+1]-[N])\psi_+\psi_-.
\end{equation}
Its spectrum is shown in the last column of Figure~3. \vskip1cm
\vbox{\begin{center} \epsfig{file=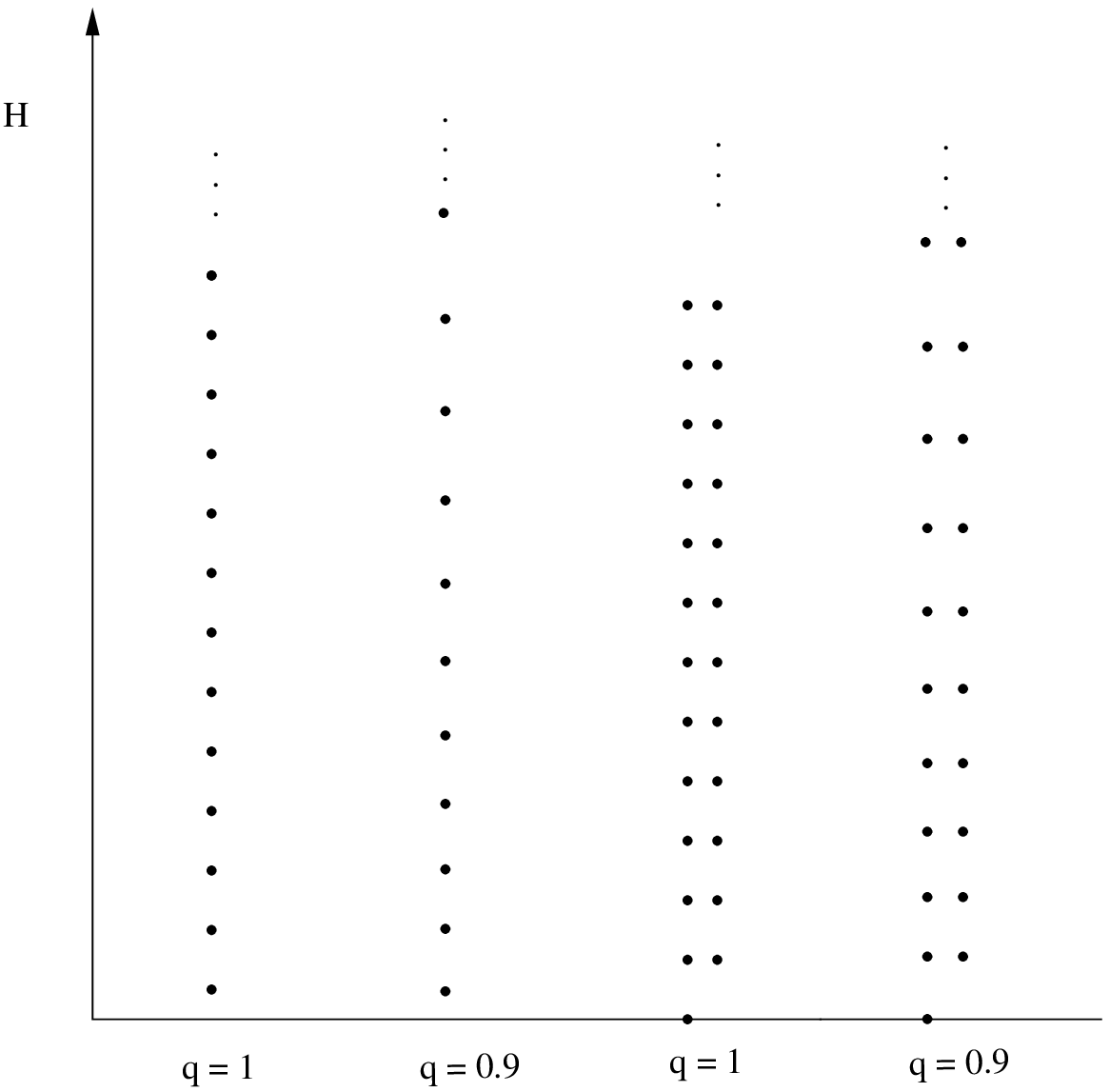}
\end{center}
\begin{center}
Figure 3. In column 4, the spectrum of the $q$-deformed SUSY h.o.
for $q=0.9$.
\end{center}}
For $q$ a root of unity, these supercharges do not lead to a
hermitian hamiltonian, so a little more work is required.
Introducing
\begin{equation}
Q_+=\sqrt{2}(a_-\psi_+),\quad\textrm{and}\quad
Q_-=\sqrt{2}(\epsilon([N])a_+\psi_-),
\end{equation}
we get a supersymmetric hermitian hamiltonian
\begin{equation}
H=\frac{1}{2}\{Q_+,Q_-\}=|[N]|+(|[N+1]|-|[N]|)\psi_+\psi_-,
\end{equation}
with two zero energy ground states and whose non-zero spectrum is
quadruply degenerate (Figure~4). There are four supercharges.
Again, there is a finite number of states, with
$E_{max}\rightarrow k/2\pi$. \vskip1cm \vbox{\begin{center}
\epsfig{file=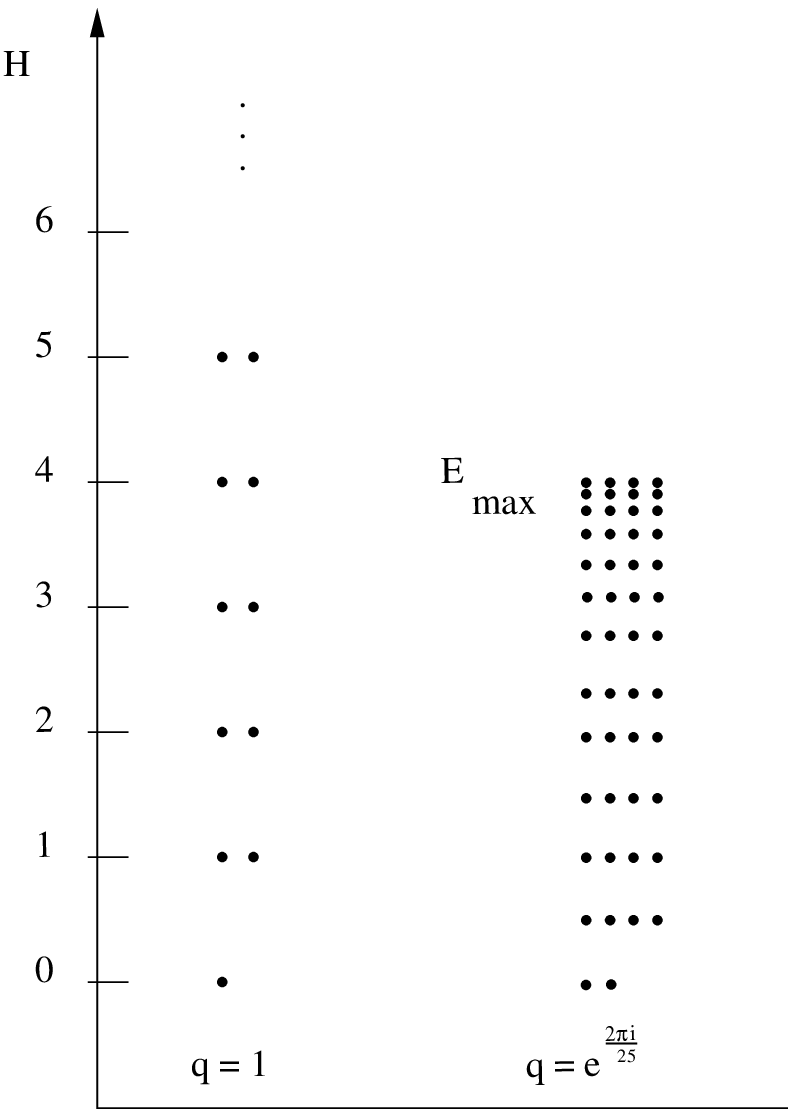}
\end{center}
\begin{center}
Figure 4. The spectra of the $q$-deformed SUSY s.h.o. for
$q=e^{2\pi i/25}$.
\end{center}}

\subsection{Supersymmetry breaking}
\begin{flushleft}
\underline{Example 1}: \end{flushleft}

 Let us now consider adding a quartic
hermitian potential to the supersymmetric hamiltonian $H_B$ of
earlier.
\begin{equation}
H = H_B + \frac{\lambda}{16}X^4.
\end{equation}
This hamiltonian is not supersymmetric. The spectrum shows the
small splittings one would expect (Figure 5). The ground state
energy is different from zero by a quantity of order $\lambda$ and
the energy splittings of the higher energy states are also of
order $\lambda$.
 \vskip1cm \vbox{\begin{center} \epsfig{file=
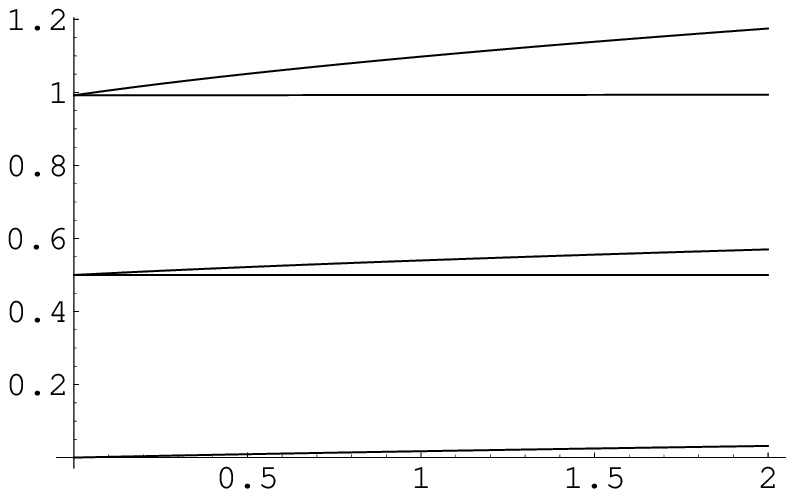}
\end{center}
\begin{center}
Figure 5. Splitting of eigenvalues by $\lambda X^4$ potential.
\end{center}}

\begin{flushleft}
\begin{underline}
{Example 2}: \end{underline}
\end{flushleft}

Consider a one-dimensional supersymmetric quantum mechanics
model~\cite{Witten:nf} with a supercharge $Q= P\psi_1 +
W(X)\psi_2$, and commutation relations $[X, P]= i$, and $\{\psi_i,
\psi_j\}=\delta_{ij}$. The hamiltonian is
\begin{equation}
H= \frac{1}{2}\left\{Q,Q\right\} = \frac{P^2}{2} + \frac{W^2}{2} +
i W' \psi_1\psi_2.
\end{equation}
We would like to consider a badly behaved superpotential, for
example $W=g/X$. Then $H=
\frac{1}{2}\left(P^2+g^2/X^2+g\sigma_3/X^2\right)$. For an odd
root of unity, the matrix $X$ is not invertible (0 is an
eigenvalue.) Thus let's deform by an even root of unity,
$q=e^{i\pi/k}$. Consider the representation with $k$ states built
as $\frac{a_+^n}{\sqrt{|[n]|}}|0\rangle$ for which $X$ does not
have a zero eigenvalue. The matrix $X$ is now invertible. We then
simulated the system on a classical computer. The spectrum has a
double degeneracy, which is irrelevant here.
\begin{center}
\begin{tabular}{r|ll|l}
$k$ &$E_0$ &$E_1$ &$E_1-E_0$ \\
\hline \\
40 & 0.404 & 0.696 & 0.292 \\
80 & 0.202 & 0.348 & 0.146 \\
160 & 0.101 & 0.174 & 0.073 \\
320 & 0.0504 & 0.0868 & 0.0364 \\
640 & 0.0252 & 0.0434 & 0.0182 \\
\end{tabular}
\end{center}
\begin{center}
Table 1: lowest two energy levels for $W= 3/X$.
\end{center}
In table 1, we have tabulated the numerical values of the lowest
two energy levels of this system, for different values of $k$ and
the specific case of the superpotential $W=3/X$. It is quite clear
that the model is becoming supersymmetric in the limit
$k\rightarrow\infty$. Supersymmetry is however broken for any
finite value of $k$. It is broken by a ``small'' amount though, as
it is quite apparent that the ground state energy $E_0$ goes to
zero as $1/k$ and that the energy splittings $E_1-E_0$ also go to
zero as $1/k$. For cosmological breaking of supersymmetry, one
would like the energy splittings to decrease slower, at a
fractional power of $k$ for example.

\section{Using quantum groups for black holes in de Sitter}
We started this project hoping that $SU(2)_q$ representations
would exhibit properties qualitatively similar to those of black
holes in de Sitter space. The key point is that black holes in de
Sitter space cannot be larger than the size of the cosmological
horizon. This raises a question: suppose one takes a black hole
which is almost the size of the cosmological horizon, and throws
in it some extra matter. What happens? We will naively assume that
we can have such matter at our disposal and that the matter can be
thrown into the black hole while keeping the size of the
cosmological horizon fixed. Then one scenario is that such a black
hole could behave like a quantum group. Namely, as we throw in
more and more matter, the black hole gets smaller and smaller, and
its energy decreases. We will illustrate how $SU(2)_q$ at a root
of unity has this curious feature. After a review of the fairly
well-known properties of $SU(2)_q$, we will give our results on
using $SU(2)_q$ for black holes.

\subsection{Review of $SU(2)_q$}

The reader familiar with $SU(2)_q$ would like to skip this
section. Quantum groups have several applications in theoretical
physics, such as the quantum inverse scattering method and
integrable models, conformal field theory in two dimensions,
nuclear physics and non-commutative geometry. For an introduction
addressed to physicists, see again Biedenharn's
book~\cite{Biedenharn:vv}.

Now we will review some well-known facts about $SU(2)_q$. This
quantum group is by definition the universal enveloping algebra
generated by $J_\pm$ and $K$ subject to the commutation relations:
\begin{equation}
KJ_\pm K^{-1} = q^{\pm 2}J_\pm \qquad [J_+,J_-] =
\frac{K-K^{-1}}{q-q^{-1}}=[2J_3].
\end{equation}
Here $q=e^{2\pi i/k}$, and $k$ is a positive odd number (the case
with $k$ even is slightly more complicated). As $q\rightarrow 1$,
these commutation relations reduce to those of the Lie algebra of
$SU(2)$, for $K=q^{2J_3}$. The hermiticity operation acts as
$K^\dagger=K^{-1}$ and $J_\pm^\dagger=J_\mp$.  The generalization
of the Casimir invariant of $SU(2)$ is:
\begin{equation}
C = J_- J_+ + \left[J_3+\frac{1}{2} \right]^2 -
\left[\frac{1}{2}\right]^2.
\end{equation}
One can easily check that $C$ commutes with $J_\pm$ and $K$ by
using the identities $[x+y] = q^{x} [y] + [x] q^{-y}$ and
$[x][y-z]+[y][z-x]+[z][x-y]=0$. We will use this Casimir invariant
as our hamiltonian (the constant $[1/2]^2$ subtracted to make the
ground state energy vanish). As we will see below, there are many
representations of this algebra, and most of them can be ignored
for our purposes. These representations are labelled by three more
Casimir invariants, $J_+^k$, $J_-^k$, $K^k$. They are not fully
independent, as there is one algebraic relation between the 4
invariants.

\subsubsection{The representations of $U_q(SU(2))$}

When $q$ is real, the representations of $U_q(SU(2))$ are a
continuous deformation of the representations of $SU(2)$. When $q$
is a root of unity, the situation is drastically different.

(A) The closest representations to those of $SU(2)$ arise when
$J_\pm^k=0$. Then we have unitary irreducible representations for
angular momentum $j= 0,1/2,1,\ldots,j_{max}$ up to a maximum value
of $j_{max} = \frac{k}{2} -1$. These representations have $2j+1$
states, and are continuous deformations of the usual
representations of $SU(2)$. The representation with $j =
\frac{k-1}{2}$ is not unitary, but still irreducible. Then, there
is a family of representations of dimension $2k$. These
representations are both non-unitary and reducible but not fully
reducible (this is called indecomposable). These non-unitary
representations are quite undesirable physically. We will be
cavalierly dropping them, but see~\cite{Mack:1992um} for the
rational behind this statement.

(B) When $J_\pm^k\neq 0$, there are cyclic representations, of
dimension $k$ and $2k$, labelled by families of continuous
parameters. Some of them are unitary and others are not. These
representations form a separate ring, i.e. separate from the
$J_\pm^k=0$ representations, and we can also ignore them for our
purposes, as we did for the cyclic representations of $q$-bosons.
In physical terms, the eigenvalues of $J_\pm^k$ label different
solitonic superselection sectors, and we declare that we work only
in the $J_\pm^k=0$ sector.

Therefore, the only physical representations are
$j=0,1/2,\ldots,j_{max}$ of (A) and, what is most essential for
us, there is a finite number of them.

Returning to the Casimir, evaluated on the unitary representations
$j=0,\ldots, j_{max}$:
\begin{equation}
C = \frac{\sin^2\frac{\pi (2j+1)}{k}- \sin^2\frac{\pi}{k}
}{\sin^2\frac{2\pi}{k}} = \frac{1}{2\sin^2\frac{2\pi}{k}}\left(
\cos\frac{2\pi}{k}-\cos\frac{2\pi(2j+1)}{k} \right).
\end{equation}
There is also a notion of $q$-dimension $D$ of a representation:
\begin{equation}
D = [2 j + 1] =
\frac{\sin\frac{2\pi(2j+1)}{k}}{\sin\frac{2\pi}{k}}.
\end{equation}
For example, if $k=5$, the $j=0,1/2,1,3/2$ representations are
unitary; $j=0,3/2$ have $q$-dimension 1 and Casimir $0$; $j=1/2,1$
have $q$-dimension $\frac{1-\sqrt 5}{2}$ and Casimir $-1+\sqrt 5$.
But the 5-dimensional $j=2$ is not unitary, its $q$-dimension is
zero. The $j=5/2$ representation is part of the 10-dimensional
$``9/2= 5/2+2"$ indecomposable representation, also with
$q$-dimension 0.

A key feature of the spectrum as given by the Casimir is a
``hump", which becomes more apparent for larger values of $k$.
Since $C$ is given by a trigonometric functions, we see that
$C(j)$ starts at zero, increases up to a maximum value $j_{top} =
\frac{k-3}{4}$ at the ``top of the hill", where it takes a maximum
value of $C_{top} \approx \frac{k^2}{4\pi^2}$ for large $k$, and
then decreases back to zero.

\subsubsection{Tensor products of representations}
The formula we will be using for decomposing the tensor product of
the unitary representations is quite simple~\cite{Biedenharn:vv}:
\begin{equation}
 \left(j_1\otimes j_2\right)= \left(
\bigoplus_{j=|j_1-j_2|}^{\min(j_1+j_2,k-j_1-j_2-2)}j \right).
\end{equation}
This formula is obtained by throwing away the non-unitary or
indecomposable representations, as justified in~\cite{Mack:1992um}
(see section 5 on page 528)~\footnote{For completeness, we will
give the formulas including these
representations~\cite{Arnaudon:1992ig}. First, when we tensor two
unitary representations of spin small enough, the decomposition is
analogous to $SU(2)$:
\begin{equation}
\left(j_1\otimes j_2\right)\Big\vert_{j_1+j_2 < \frac{k-1}{2}}  =
\bigoplus_{j=|j_1-j_2|}^{j_1+j_2}j.\nonumber
\end{equation}
When the sum of the spins of the two unitary representations we
multiply is larger than $\frac{k-1}{2}$, then
non-unitary/indecomposable representations occur in the
decomposition:
\begin{eqnarray}
 \left(j_1\otimes j_2\right)\Big\vert_{ 2(j_1+j_2)\ {\textrm even}, \ge k-1} & = &\left(
\bigoplus_{j=|j_1-j_2|}^{\min(j_1+j_2,k-j_1-j_2-2)}j \right)
\oplus \frac{k-1}{2} \oplus \left(m_{j_1,j_2} \bigoplus \frac{2k-1}{2}\right) \nonumber\\
\left(j_1\otimes j_2\right)\Big\vert_{ 2(j_1+j_2)\ {\textrm odd},
\ge k-1} & = &
\left(\bigoplus_{j=|j_1-j_2|}^{\min(j_1+j_2,k-j_1-j_2-2)}j
\right)\oplus  \left(m_{j_1,j_2} \bigoplus \frac{2k-1}{2}\right),
\nonumber
\end{eqnarray}
where $ m_{j_1,j_2}$ is the known number of times that the various
$k-\frac{1}{2}$ representations occur, which can be worked out by
equating the total dimensions on both sides.}.

Let's give two examples of the use of the formula: for a small
number like $k=5$ and a larger number like $k=103$. For $k=5$, the
tensor product of two spins $\frac{1}{2}$ is as usual: $1/2\otimes
1/2 = 0+1$, but other tensor products are truncated:  $1/2\otimes
3/2 = 1$, $1\otimes 1 = 0 + 1$, $1\otimes 3/2 = 1/2$, $3/2\otimes
3/2= 0$.

Let's see the pattern for these tensor decompositions with
$k=103$. The unitary reps are $j=0,1/2,1,\ldots,50,
\frac{101}{2}$. Then there is the non-unitary rep $j=51$ and the
indecomposable rep $j=\frac{205}{2}$. The top of the hill is at
$j_{top} = 25$. For simplicity, let's just look at tensor products
where $j$ is an integer. We get for the product of a small
representation times one near the top of the hill: $1\otimes 24 =
23\oplus 24\oplus 25$; for two very large representations: $50
\otimes 50 = 0\oplus 1$; for the tensor product of many small
representations: $1^{\otimes 50} = 50 \oplus \mu 49 \oplus
\cdots$, where the multiplicity of the $49$ appearing in the
decomposition is $\mu=49$; etc.

\subsection{Black holes in de Sitter and $SU(2)_q$}
Now, we want to use $SU(2)_q$ to make an analogy with black holes
in de Sitter. Choosing a value of $k$ is like choosing the
dimension of the Hilbert space and the size of the cosmological
horizon, and so will be very large. An elementary particle will be
a representation with $j$ small. A black hole will be a
``medium-size" unitary representation, with $j$ smaller than but
close to $j_{top} = \frac{k-3}{4}$. The energy, i.e. the size, of
black holes grows like $j$, until it abruptly stops growing when
the maximum energy/size has been reached. The ``large"
representations with $j$ larger than $j_{top}$ will form a
``hidden sector", because as we saw from the tensor product
decompositions, they are hard to reach from the point of view of
the fundamental particles in the sense that it takes of order $k$
tensor products of fundamental particles to reach them. It maybe
be that the universe started in a state constituting mostly of
small $j$ particles. For $k$ large, it may be that the time scale
for populating the hidden sector is very large compared to the
present age of the universe.

Now, when we ``drop" a fundamental particle into a black hole, we
use the tensor product decomposition to find the result of the
interaction. What we get are black hole states, some of them
conventional with $j<j_{top}$, but others have $j>j_{top}$ and
thus have lower energy: dropping a particle into the black hole
has had the effect of ``shrinking" some of the black hole states,
lowering their energy, and putting them into the ``hidden sector".

When two black holes interact, states with all values of $j$ are
produced. When two states from the hidden sector interact, they
collapse mostly into fundamental particles. Clearly, one would
need a theory of the dynamics to make sense of these interactions.

In summary, the merit of this caricature of de Sitter space is
that black holes have a maximal size and energy. When more matter
is dropped into such a black hole, it manages to shrink and lower
its energy.

\subsection{The thermodynamics for $SU(2)_q$ particles, versus $SU(2)$}
Continuing the analogy, we would like to see if the natural
thermodynamics properties of $SU(2)_q$ will be useful to describe
black holes. The total number of black hole states is of order
$k^2$, much larger than the number of fundamental particles, of
order 1. For the real world, we would like a much faster,
exponential growth. Can this be obtained from quantum groups? We
will not address this question in this paper, but only begin the
discussion by looking at the high and low temperature properties
of the partition function of $SU(2)_q$.

To calculate thermodynamics properties, we should decide what we
are going to do with the states in the ``hidden sector''. One
possibility is to simply include them in the partition
function~\footnote{Another possibility, if the temperature is low
enough, is to ignore the hidden sector states and work with a
system in quasi-equilibrium. At low temperature then, rather
trivially, the partition function for $SU(2)_q$ has the same
leading behavior as the one for $SU(2)$, except that the
temperature is shifted: $Z\rightarrow 1 + 2 e^{-{\frac{3\beta}{
4}}(1-\frac{1}{3}\tan^2({\frac{\pi}{k}}))}$, that is
$\beta\rightarrow \beta (1-{\frac{1}{3}}\tan^2(\frac{\pi}{k}))$,
which goes over to the $SU(2)$ partition function for large $k$.
At high temperature, we should certainly not ignore the hidden
sector.}.

For the Casimir $C$ as the hamiltonian, with $\beta=1/T$ the
inverse temperature, the partition function is:
\begin{equation}
Z_{SU(2)_q} = \sum_{m=1}^{k-1} m e^{-{\frac{\beta}
{2\sin^2{\frac{2\pi}{
k}}}}\left(\cos{\frac{2\pi}{k}}-\cos{\frac{2\pi m}{k}}\right)}
\label{zsu2q}.
\end{equation}
It can be evaluated in close form:
\begin{align}
Z_{SU(2)_q} = k \sum_{m=1}^{k-1}  e^{-{\frac{\beta}
{2\sin^2{\frac{2\pi}{
k}}}}\left(\cos{\frac{2\pi}{k}}-\cos{\frac{2\pi m}{k}}\right)} \\
 = k e^{-\gamma \cos\frac{2\pi}{k}}\left(e^{-\gamma}+k
 I_0(\gamma)\right)
\end{align}
where $\gamma=\frac{\beta}{2\sin^2\frac{2\pi}{k}}$ and $I_0$ is a
Bessel function. For undeformed $SU(2)$, with the Hamiltonian
$H=j(j+1)$, the partition function is:
\begin{equation}
Z_{SU(2)}  =  \sum_{n=0}^{\infty}(n+1) e^{-\beta  n ( n + 2 )/4}.
\end{equation}
The usual thermodynamics quantities are the free energy $F$
defined by $Z= e^{-\beta F}$, the entropy $S= \beta^2 dF/d\beta$,
the average energy $E = d(\beta F)/d\beta$ with $E=F+TS$, and the
specific heat $c_v = dE/dT$.

\subsubsection{At low temperature}

For $SU(2)$ at low temperature, the leading contribution to the
partition function of course comes from the lowest energy states,
so we have $Z\rightarrow 1+ 2 e^{-3 \beta/4}$, $F\rightarrow
\frac{-2}{\beta} e^{-3\beta/4}$, $S\rightarrow
(2+3\beta/2)e^{-3\beta/4}$, $E\rightarrow \frac{3}{2}
e^{-3\beta/4}$ and $c_v\rightarrow \frac{9\beta^2}{8}
e^{-3\beta/4}$. The specific heat is of course very small at low
temperature.

For $SU(2)_q$, we have $Z\rightarrow k (1+e^{-\beta
(1-\frac{1}{4}\cos^{-2}\frac{\pi}{k})})$. This partition function
does not at all reduce to the one for $SU(2)$ in the $k\rightarrow
\infty$ limit. This is because of the low energy states from the
hidden sector. Instead, as $k\rightarrow\infty$, we have
$Z\rightarrow k+ke^{-3\beta/4}$, from which we get $F\rightarrow
\frac{-1}{\beta} (\log k + e^{-3\beta/4})$, $E\rightarrow
\frac{3}{4} e^{-3\beta/4}$, $S\rightarrow \log k +
(1+3\beta/4)e^{-3\beta/4}$ and $c_v\rightarrow \frac{9\beta^2}{16}
e^{-3\beta/4}$.

\subsubsection{At high temperature}

The partition function of $SU(2)$ at high temperature goes like
$Z_{SU(2)}\rightarrow \frac{2}{\beta}$. The free energy is
negative and goes as $\frac{1}{\beta}\log{\frac{\beta}{2}}$,
characteristic of one quantum mechanical degree of freedom. The
entropy is $S= -\log \frac{\beta}{2}+1$, the energy $E =
\frac{1}{\beta}$ and $c_v = 1$.

At high temperature and large $k$, the partition function for
$SU(2)_q$ becomes $Z\rightarrow k^2(1-\frac{\beta k^2}{8 \pi^2})$,
valid when $\beta k^2$ is small. From that we get: $F= -\frac{\log
k^2}{\beta} + \frac{k^2}{8 \pi^2}$. The free energy and entropy
are characteristic of a quantum mechanical system with $k^2$
states. $E= \frac{k^2}{8\pi^2}$; $S= \log k^2$ and $c_v=0$. We see
that raising the temperature does not further increase the
entropy. The average energy is half the maximum energy level:
further increasing the temperature does not increase the average
energy and we never reach $E_{max}$.

Clearly, the high temperature properties of $SU(2)_q$ are typical
of a system with a finite number of states. There is a maximum
average energy, a maximum entropy and the specific heat goes to
zero. These are desirable features for the thermodynamics of de
Sitter space.


\section{A toy model for de Sitter space (without quantum groups)}

 In this section, we will
describe a simple quantum mechanical model with only two degrees
of freedom that very roughly features a property that we expect a
complete quantum theory of gravity to possess: that there is a
finite number of states for a positive cosmological constant, and
an infinite number of states for a negative one.

The authors of~\cite{Bellucci:2001xp} made the following very
interesting observation. A noncommutative quantum mechanics model
on the infinite plane, with a magnetic field, with two degrees of
freedom, has a conserved angular momentum $J$. The model has a
parameter, let us call it $\Lambda$ in analogy with the
cosmological constant. When this parameter is negative or zero,
there is an infinite number of quantum states, as one would expect
for anti-de Sitter space and Minkowski space. When the parameter
is positive, there is a finite number of states, as one would
expect for de Sitter space, as long as one considers only the
sector with a given value of the conserved total angular momentum.

In section 4.1, we will review the model
of~\cite{Bellucci:2001xp}. In section 4.2, we look at what happens
to the BNS model if we deform it by a root of unity. In section
4.3, we supersymmetrize the BNS model. We looked at breaking
supersymmetry by a large amount, but did not find a way to do it.

\subsection{The model of Bellucci, Nersessian and Sochichiu}

The model~\cite{Bellucci:2001xp} is a 2-dimensional noncommutative
quantum mechanics system with a magnetic field. The hamiltonian is
$H= \frac{p^2}{2} + \frac{\omega^2 x^2}{2}$, with the commutation
relations $[x_1,x_2] = i\theta$, $[x_i,p_j]=i\delta_{ij}$ and
$[p_1,p_2]=iB$, and $\theta>0$. Since the hamiltonian is
quadratic, it can be diagonalized exactly by a Bogolyubov
transformation~\cite{Nair:2001ii}. The result is
\begin{equation}
H_1= \frac{\omega_+}{2}(\beta_+\beta_-+\beta_-\beta_+)+
\frac{\omega_-}{2}(\alpha_+\alpha_-+\alpha_-\alpha_+)
\label{ham1},
\end{equation}
with $[\alpha_-,\alpha_+] =1$, $[\beta_-,\beta_+] = \sgn
(\Lambda)$ and $[\alpha,\beta]=0$. Thus the spectrum is the
product of two decoupled harmonic oscillators with different
frequencies, $H=\omega_+
(n_++\frac{1}{2})+\omega_-(n_-+\frac{1}{2})$, with $n_\pm =
0,1,2,\ldots$ and
\begin{equation}
\omega_\pm=\frac{1}{2}\Big|B+\omega^2\theta\pm\sqrt{(B-\omega^2\theta)^2+4\omega^2}\Big|=
\frac{1}{4\theta}\Big|{\cal E}+\Lambda\pm \sqrt{({\cal E} -
\Lambda)^2+ 4\Lambda}\Big|, \label{frequencies}
\end{equation}
for ${\cal E} = 1+
\omega^2\theta^2$. The physics is quite different depending on the
sign of $\Lambda=\theta B-1$. There is a conserved angular
momentum $j = n_+ + \sgn(\Lambda)\ n_-$. At fixed values of the
angular momentum, if $\Lambda$ is negative, there is an infinite
number of states allowed, but if $\Lambda$ is positive, there is
only a finite number of states.

The transformation required to get to this hamiltonian depends on
the sign of $\Lambda$~\cite{Bellucci:2001xp}. At a first stage,
one goes from the variables $(x,p)$ to an intermediate set
$(a,b)$. At this stage the hamiltonian takes the form:
\begin{equation}
H_2 =
\frac{|\Lambda|}{2\theta}(b_+b_-+b_-b_+)-\frac{i\sqrt{|\Lambda|}}{\theta}
(b_+a_--a_+b_-)+ \frac{{\cal E}}{2\theta}(a_+a_-+a_-a_+),
\label{ham2}
\end{equation}
where $a_\pm= \frac{x_1\mp i x_2}{\sqrt{2\theta}}$, $b_\pm=
\sqrt{\frac{\theta}{2|\Lambda|}}(\pi_1\mp i\pi_2)$ and $\pi_i =
p_i -\epsilon_{ij} x_j/\theta$.

At the second stage, then for $\Lambda <0$, the operators $\alpha$
and $\beta$ are related to $a$ and $b$ by the further
substitutions: $\binom{a_+}{b_+}= U\binom{\alpha_+}{\beta_+}$,
with $U=
\left(\begin{matrix}\phi\cosh\chi & \phi\sinh\chi \\
\bar\phi\sinh\chi & \bar\phi\cosh\chi \end{matrix}\right)$, $\tanh
2\chi = \frac{2\sqrt{|\Lambda|}}{ {\cal E}-\Lambda}$, and
$\phi=e^{i\pi/4}$.

On the other hand, when $\Lambda >0$, the change of variable is
$\binom{a_+}{b_+}= U\binom{\alpha_+}{\beta_+}$ with $U=
\left(\begin{matrix}\phi\cos\chi & \phi\sin\chi \\
-\bar\phi\sin\chi & \bar\phi\cos\chi \end{matrix}\right)$, $\tan
2\chi = \frac{2\sqrt{\Lambda}}{ {\cal E}-\Lambda}$.

\subsection{$q$-deformation with 2 degrees of freedom}

\begin{flushleft}
\begin{underline}
{Deformation 1}: \end{underline}
\end{flushleft}
The most straightforward deformation is to deform the oscillators
$\alpha$, $\beta$, at the stage where the hamiltonian is diagonal,
in the form~\ref{ham1}.

One can choose to deform by a root of unity when $\Lambda>0$:
\begin{align}
\alpha_- \alpha_+ - q \alpha_+  \alpha_-= L^{-1}_\alpha, \qquad
L_\alpha\alpha_{\pm} L^{-1}_\alpha=q^{\pm 1} \alpha_{\pm}\\
\beta_-\beta_+ - q \beta_+\beta_- = L^{-1}_{\beta}, \qquad
L_{\beta}  \beta_\pm L^{-1}_{\beta} = q^{\pm 1} \beta_\pm,
\label{deform1}
\end{align}
with $[\alpha,\beta]=0$, and pick a hamiltonian, for example we
can take $H_B$ from section 2.3 to get a supersymmetric model:
\begin{equation}
H= \frac{\omega_+}{2}\Big|\beta_+\beta_-+\beta_-\beta_+\Big|+
\frac{\omega_-}{2}\Big|\alpha_+\alpha_-+\alpha_-\alpha_+\Big|,
\end{equation}
while one can deform by $q$ real when $\Lambda<0$:
\begin{align}
\alpha_- \alpha_+ - q \alpha_+  \alpha_-= q^{-N_\alpha}, \qquad
[N_\alpha,\alpha_{\pm}]=\pm \alpha_{\pm}\\
\beta_+\beta_- - q \beta_-\beta_+ = q^{-N_\beta}, \qquad [N_\beta,
\beta_\pm] = \mp \beta_\pm, \label{deform2}
\end{align}
while again $[\alpha,\beta]=0$ and note that
$\beta_+\leftrightarrow\beta_-$ have exchanged roles. One could
imagine a formula such as $q= e^{i \sqrt{\Lambda}}$, which
requires $\Lambda$ to be quantized if it is
positive~\cite{Major:1995yz}, while if $\Lambda$ is negative, $q$
is real. This deformation removes the requirement of fixing the
value of the angular momentum $J$ to get a finite number of states
for $\Lambda>0$.

\begin{flushleft}
\begin{underline}
{Deformation 2}: \end{underline}
\end{flushleft}
We could deform the model at an earlier stage, when the
hamiltonian is not yet diagonal, but the commutation relations are
canonical~\footnote{We do not know how to deform the model when
the commutation relations are not canonical.}. When the
hamiltonian takes the form~\ref{ham2}, we can deform the
oscillators $a$ and $b$. Depending on the sign of $\Lambda$, we
would use
\begin{align}
a_- a_+ - q a_+  a_-= L^{-1}_a, \qquad
L_a a_{\pm} L^{-1}_a=q^{\pm 1} a_{\pm}\\
b_-b_+ - q b_+b_- = L^{-1}_b, \qquad L_b b_\pm L^{-1}_b = q^{\pm
1} b_\pm,
\end{align}
for $\Lambda>0$, with $[a,b]=0$ or
\begin{align}
a_- a_+ - q a_+ a_-= q^{-N_a}, \qquad
[N_a,a_{\pm}]=\pm a_{\pm}\\
b_+b_- - q b_-b_+ = q^{-N_b}, \qquad [N_b,b_\pm] = \mp b_\pm,
\end{align}
with $\Lambda<0$ and again $[a,b]=0$.

Since the hamiltonian is no longer diagonalizable exactly, we
studied this deformation with Mathematica. We include a sample of
the code in an appendix, to illustrate the straightforwardness of
doing such numerical computations. In Deformation 1, for $\omega_+
\gg \omega_-$, the spectrum exhibits clear Landau levels. The
simulation indicates that Landau level features are still somewhat
present for Deformation 2.

\subsection{Noncommutative SUSY quantum
mechanics with a magnetic field}

We have seen previously that the model of BNS features a
non-analyticity in $\Lambda$: the hamiltonian depends on
$|\Lambda|$, the harmonic oscillator frequencies $\omega_\pm$
feature an absolute value, and the spectrum is qualitatively
different depending on the sign of $\Lambda$. We thus decided to
supersymmetrize this model to see if it would feature some
enhancement of supersymmetry breaking as desired in a cosmological
supersymmetry breaking scenario. We would need for that to find a
quantity that depends on the derivative of $\Lambda$. The
derivatives of the energy are discontinuous. For example, the
ground state energy:
\begin{equation}
\frac{\partial E}{\partial \Lambda}\Big\vert_{\Lambda\geq 0} = 1 \
\quad {\textrm {but}}\ \qquad \frac{\partial E}{\partial
\Lambda}\Big\vert_{\Lambda\leq 0} =
\frac{1-\theta^2\omega^2}{1+\theta^2\omega^2}\rightarrow -1\
\textrm {for\ large} \ \theta\omega.
\end{equation}
There are two bosonic degrees of freedom $X_i$, $P_i$, $i=1,2$,
and four real fermions $\psi_\alpha$, $\alpha=1,\ldots,4$. They
satisfy the commutation relations:
\begin{equation}
[X_1,X_2]=i\theta,\quad [X_i,P_j]=i\delta_{ij},\quad
[P_1,P_2]=iB,\quad\textrm{and}\quad
\{\psi_\alpha,\psi_\beta\}=\delta_{\alpha\beta}.
\end{equation}
It is more convenient to work with complex/light-cone coordinates:
\begin{equation}
X_\pm = \frac{X_1\mp i X_2}{\sqrt 2}, \quad P_\pm=\frac{P_1\mp i
P_2}{\sqrt 2}, \quad \psi_\pm = \frac{\psi_1\pm i\psi_2}{\sqrt 2},
\quad\textrm{and}\quad \chi_\pm = \frac{\psi_3\pm i \psi_4}{\sqrt
2}.
\end{equation}
The commutators are now:
\begin{equation*}
[X_-,X_+]=\theta,\quad [X_\pm,P_\mp]=i,\quad [P_-,P_+]=B, \quad
\{\psi_+,\psi_-\}=1,\quad \textrm{and} \quad \{\chi_+,\chi_-\}=1.
\end{equation*}
There are two supercharges:
\begin{equation}
Q_+= \sqrt{2}\left( P_- \psi_+ + W(X_-) \chi_+\
\right),\quad\textrm{and}\quad Q_-= \sqrt{2}\left( P_+ \psi_- +
\overline{W}(X_+) \chi_-\ \right),
\end{equation}
from which we get the hamiltonian:
\begin{align}
H&=\frac{1}{2}\{Q_+,Q_-\}=\left(P_+P_-+ \overline{W}(X_+)W(X_-)\right)+ \nonumber \\
&\left(\theta
W'(X_-)\overline{W}'(X_+)\chi_+\chi_-+B\psi_+\psi_-+iW'(X_-)\chi_+\psi_--i\overline{W}'(X_+)\psi_+\chi_-
\right).
\end{align}
>From now on, we will specialize to the case of a linear
superpotential $W(X_-)=\omega X_-$, and then the hamiltonian is
free:
\begin{equation}
H=P_+P_-+\omega^2
X_+X_-+\theta\omega^2\chi_+\chi_-+B\psi_+\psi_-+i\omega\chi_+\psi_--i\omega\psi_+\chi_-.
\end{equation}
The bosonic part being the same as for the BNS model, except for
the shift in ground state energy, we will just quote the formulas
for diagonalizing the fermions. The fermionic part of $H$ takes
the form
\begin{equation}
\begin{pmatrix}
 \chi_+ & \psi_+ \\
\end{pmatrix}
\begin{pmatrix}
 \theta\omega^2 & i\omega \\
-i\omega &  B  \\
\end{pmatrix}
\begin{pmatrix}
\chi_- \\ \psi_- \\
\end{pmatrix}.
\end{equation}
Its eigenvalues are $\omega_\pm$. The eigenvectors are, for
$\Lambda>0$:
\begin{equation}
v_\pm = ((B-\theta\omega^2)^2+4\omega^2)^{-1/4}\begin{pmatrix}i
(\omega_\pm+B)^{1/2}
 \\ \omega(\omega_\pm+B)^{-1/2}
\end{pmatrix}.
\end{equation}
 After
diagonalization, the hamiltonian takes the form:
\begin{equation}
H=\omega_+(n_++\frac{1}{2}\pm\frac{1}{2})+\omega_-(n_-+\frac{1}{2}\pm\frac{1}{2}),
\end{equation}
with the frequencies given by the same
expression~\ref{frequencies} as in the BNS model.

Next, one can $q$-deform the model, along Deformation 1 or 2 above
for example, and analyze the spectrum. Again, we find
supersymmetry breaking by a small amount, qualitatively similar to
the results presented in Table 1.

\section{Conclusion}
To get a parametrically large breaking of supersymmetry, one would
hope that if some of the matrix elements in the hamiltonian differ
by some small amount of order $\epsilon$, then the splitting in
eigenvalues is enhanced to order $\epsilon^\alpha$ for $\alpha\in
(0,1)$. For example, this would arise for the hamiltonian
$\bigl(\begin{smallmatrix}1+\epsilon&-1\\ 1&
-1\end{smallmatrix}\bigr)$, for which the eigenvalue splitting is
of order $\sqrt\epsilon$. However this hamiltonian is not
hermitian. On a highly speculative note, it would be very
interesting if unitarity violation was the cause of supersymmetry
breaking, and of the cosmological constant.

\acknowledgments

We would like to thank Willy Fischler for suggesting this project,
and for encouragements and discussions at all stages of this
project. We would also like to acknowledge useful correspondence
with V.~Nair, A.~Nersessian, M. Plyushchay, Y. Saint-Aubin and
C.~Sochichiu and thank S. Bellucci, D. Freed, J. Distler, S.
Majid, E. Nicholson, S. Paban and L. Smolin for comments,
criticisms or discussions. This work was supported in part by NSF
grant PHY-0071512 and in part by a PPARC post-doctoral grant.

\appendix

\section{Mathematica Code}
This very simple code computes the spectrum of the hamiltonian of
section 4.2. It should be easy to see how to adapt it to perform
the other computations presented in this article.

\begin{flushleft}
{\texttt{$<$$<$ `LinearAlgebra`MatrixManipulation}

\texttt{k = 5;}

\texttt{q = Exp[2 Pi I/k];}

\texttt{n[x\_] := (q\^{}x - q\^{}(-x))/(q - q\^{}(-1));}

\texttt{r[x\_] := (Abs[n[x]])\^{}(1/2); }

\texttt{epsilon[x\_] := If[x $<$ 0, -1, 1];}

\texttt{aplus = Table[Switch[i-j, 1, r[j+1], \_, 0], {i, 0, k-1},
{j, 0, k-1}];}

\texttt{aminus =
    Table[Switch[i-j, -1, epsilon[n[j]]*r[j], \_, 0], {i, 0, k-1}, {j, 0, k-1}];}

\texttt{absaminus = Abs[aminus];}

\texttt{d1 = DiagonalMatrix[Table[1, {i, 1, k}]];}

\texttt{ap1 = BlockMatrix[Outer[Times, d1, aplus]];}

\texttt{am1 = BlockMatrix[Outer[Times, d1, absaminus]];}

\texttt{bp1 = BlockMatrix[Outer[Times, aplus, d1]];}

\texttt{bm1 = BlockMatrix[Outer[Times, absaminus, d1]];}

\texttt{hamiltonian[lambda\_, calE\_] := (Abs[lambda]/2)(bp1 . bm1
+ bm1. bp1) -}

\texttt{-I(Abs[lambda])\^{}(1/2)(bp1.am1-ap1.bm1) +
(calE/2)(ap1.am1+am1.ap1);}

\texttt{eig[lambda\_, calE\_] :=
Eigenvalues[N[hamiltonian[lambda,calE]]]; }}
\end{flushleft}

\vfill\eject

\end{document}